\documentclass[manuscript]{aastex}

\newcommand{\crra}{Cram\'er-Rao}

\slugcomment{Accepted by PASP}

\shorttitle{The Cram\'er \-Rao bound in astrometry}
\shortauthors{Mendez et al.}

\begin{document}

\title{Analysis and interpretation of the \crra\ lower-bound in
  astrometry: One dimensional case}

\author{Rene A. Mendez}

\affil{Departamento de Astronom\'{\i}a, Facultad de Ciencias
  F\'{\i}sicas y Matem\'aticas,\\Universidad de Chile, Casilla 36-D,
  Santiago, Chile} \email{rmendez@u.uchile.cl}

\and

\author{Jorge F. Silva and Rodrigo Lobos} \affil{Departamento de
  Ingenier\'{\i}a El\'ectrica, Facultad de Ciencias F\'{\i}sicas y
  Matem\'aticas,\\Universidad de Chile, Beauchef 850, Santiago, Chile}
\email{josilva, rlobos@ing.uchile.cl}

\begin{abstract}
In this paper we explore the maximum precision attainable in the
location of a point source imaged by a pixel array detector in the
presence of a background, as a function of the detector
properties. For this we use a well-known result from parametric
estimation theory, the so-called \crra\ lower bound. We develop the
expressions in the 1-dimensional case of a linear array detector in
which the only unknown parameter is the source position. If the object
is oversampled by the detector, analytical expressions can be obtained
for the \crra\ limit that can be readily used to estimate the limiting
precision of an imaging system, and which are very useful for
experimental (detector) design, observational planning, or performance
estimation of data analysis software: In particular, we demonstrate
that for background-dominated sources, the maximum astrometric
precision goes as $B/F^2$, where $B$ is the background in one pixel,
and $F$ is the total flux of the source, while when the background is
negligible, this precision goes as $F^{-1}$. We also explore the
dependency of the astrometric precision on: (1) the size of the source
(as imaged by the detector), (2) the pixel detector size, and (3) the
effect of source de-centering. Putting these results into context, the
theoretical \crra\ lower bound is compared to both ground- as well as
spaced-based astrometric results, indicating that current techniques
approach this limit very closely. It is furthermore demonstrated that
practical astrometric estimators like maximum likelihood or
least-squares techniques can not formally reach the \crra\ bound, but
that they approach this limit in the 1-dimensional case very tightly,
for a wide range of $S/N$ of the source. Our results indicate that we
have found in the \crra\ lower variance bound a very powerful
astrometric ``benchmark'' estimator concerning the maximum expected
positional precision for a point source, given a prescription for the
source, the background, the detector characteristics, and the
detection process.
\end{abstract}


\section{Introduction}

Astrometry relies on the precise determination of the relative location
of, usually, point sources. The estimation of the precision with which
these measurements can be done, both from an empirical, as well as
from a theoretical point of view, has been the subject of various
papers. Seminal work, as applied to stellar images recorded on
photographic plates, are those of \citet{van75} and \citet{auevan78},
with further refinements by \citet{leevan83}, in which statistical
estimations for the precision of the position of stellar images were
compared to the results from the actual fitting of stellar profiles
measured using microdensitometer scans, through a classical least
squares minimization technique
assuming a Gaussian noise on the measured intensities.

Nowadays, discrete digital detectors, such as Charged Coupled Devices
(CCDs, \citet{how06}), being highly efficient area detectors, are
widely used in astronomy for photometric, astrometric and
spectroscopic observations (\citet{mac86}, for the specific use of
CCDs in astrometry see e.g. \citet{mon92}, \citet{lin05}, and
\citet{how13}). This prompted \citet{kin83} to carry out a similar
analysis for CCDs, 
specifically for HST data, starting also from the
assumption that a least squares minimization approach provides the
best estimation of the relevant parameters.




The studies by \citet{leevan83} and \citet{kin83} (see also
\citet{sto89}) provide estimates of the statistical uncertainties of
the fitted parameters, given a noise model for the data. However, a
related question, less often addressed, is what would be the {\it
  maximum attainable precision} with which one could expect to
estimate the astrometric position of a source, given a prescription of
the detection process. This question constitutes a central aspect to
astrometric work. For example, in situations, when the detector nearly
critically samples the light distribution for point sources rendered
by the telescope optics (case, e.g., of the HST-WFPC
imager\footnote{Instrument handbook available at
  http://www.stsci.edu/hst/wfpc2/documents/handbook/IHB\_17.html, last
  accessed on December 2012.}), the question may arise on how well in
principle the flux and the position of a point source located on some
unknown background may be jointly estimated. The answer may be needed
in instrumental design, for planning observation campaigns, or for
checking the quality of data analysis algorithms.

In this paper, we concentrate, precisely, on deriving a lower bound to
the expected astrometric error for the position of a source in one
dimension (hereafter referred to as 1-D), as specified by the variance
of the position itself, for the kind of data expected from
astronomical CCD-observations. Some seminal work in this area, using
the so-called \crra\ bound (\citet{ra45}, \citet{cr46}), has been
presented by \citet{lin78}, \citet{ja92}, \citet{za95}, \citet{ad96},
\citet{lin97} and \citet{ba04}.
More recently, \citet{lin10} has published a review paper exploring
the astrometric bounds, using some of the \crra\ prescriptions, but
focusing mostly on space-based near-diffraction limited imaging. A
particularly relevant and inspirational early work on the subject is
that contained in \citet{win86} which was focused, however, on
read-out noise limited devices and applications. We use Winick's paper
as a starting point of our discussion, expanding it to sky-background
limited astronomical observations, and develop it further to explore
the general \crra\ bound for astrometry in the 1-D case. We also
elaborate on some useful approximations that explicitly expose the
dependency of the expected minimum astrometric uncertainty as a
function of the parameters characterizing the source and the
detector. Some interesting open astrophysical problems that require
high-accuracy astrometric measurements have been recently described in
the book by \citet{van13}. One may use the maximum expected accuracy
estimates developed in this paper to determine whether or not a
particular astrophysical problem - and its associated accuracy
requirements - may or may not be tackled with certain instrumental
set-up and observing conditions.

\section{Preliminaries} \label{sec:pre}

To give a formal context, in this section we introduce the basic
setting of parameter estimation, as well as some related concepts and
definitions, that will be used throughout the paper.

\subsection{Parameter estimation and the \crra\ minimum variance bound} \label{subsec:parameter_est}

The parameter estimation problem at hand can be presented in
one-dimension, in general terms, as follows: Let us consider a
collection of $I_i$ (with $i=1...n$) independent and identically
distributed realizations of a random variable. In this setting, it
will be assumed that the $\{I_i:i=1,...,n\}$ measurements follow an
underlying probability (density - if continuous, or mass - if
countable) function, denoted by $f_\theta$, which depends upon a
certain target (unknown) parameter $\theta$. Then, the parameter
estimation problem reduces to find a prescription (or statistics) such
that the function $\hat{\theta}(I_1,...,I_n)$ is a good approximation
of the underlying parameter $\theta$ that generated the data.

A standard criterion adopted in statistics to estimate $\theta$ is to
consider the rule of minimum variance (denoted by $Var$), given by:

\begin{eqnarray}\label{minvar}
\hat{\theta}() & \equiv & \arg \min_{\hat\theta} Var
(\hat\theta(I_1,...,I_n)) \nonumber \\
& = & \arg \min_{\hat\theta}  \mathbb{E}_{I_1,...,I_n \sim
  f_{\theta}^n} \left( \hat \theta (I_1,...,I_n) - \theta \right)^2
\end{eqnarray}

where $\mathbb{E}$ is the expectation value of the argument and ``$
\arg \min$'' represents the argument that minimizes the
expression. Note that in the last equality we have assumed that $\hat
\theta()$ is an unbiased estimator of the parameter (i.e., that
$\mathbb{E}(\hat\theta)= \theta$), so that under this rule we are
implicitly minimizing the mean square error of the estimate with
respect to the hidden true parameter $\theta$.

Unfortunately, the general solution of equation~(\ref{minvar}) is
intractable, as in principle it requires the knowledge of $\theta$,
which is the essence of the inference problem. However, there are
performance bounds that characterize how far can we be from the
theoretical solution in equation~(\ref{minvar}), and even scenarios
where the optimal solution can be achieved in a closed-form.  One of
the most significant results in this field is the \crra\ minimum
variance bound (\citet{ra45}, \citet{cr46}) explained below.

Let $\hat \theta()$ be an unbiased estimator of $\theta$. If we define
the function $L(I_1,...,I_n;\theta)$ as the likelihood of the
observation given the model parameter $\theta$, and we have $n$
independent random variables $I_i$ driven by the probability function
$f_\theta$, then the \crra\ bound states that:

\begin{equation}\label{varcr}
Var (\hat\theta(I_1,...,I_n)) \equiv \mathbb{E}_{I_1,...,I_n\sim
  f_{\theta}^n} \left(\hat \theta(I_1,...,I_n) - \theta \right)^2 \geq
\frac{1}{ \mathcal{I}_{\theta}(n) }
\end{equation}

provided that we satisfy the constraint:

\begin{equation}\label{constr}
\mathbb{E}_{I_1,...,I_n\sim f_{\theta}^n} \left(
\frac{d}{d \theta} \ln L(I_1,...,I_n;\theta) \right) =0
\end{equation}

and where:

\begin{equation} \label{fisher}
\mathcal{I}_{\theta}(n) \equiv \mathbb{E}_{I_1,...,I_n\sim
  f_{\theta}^n} \left( \left( \frac{d }{d \theta} \ln
L(I_1,...,I_n;\theta) \right) ^2 \right)
\end{equation}

is called the {\em Fisher information} of the data about $\theta$.

A powerful corollary of this result is that the minimum variance of
any unbiased estimator that satisfies equation~(\ref{constr})
is always going to be greater than the pre-specified quantity given by
equation~(\ref{varcr}).


Generally, this bound is not attained for the minimum variance
estimator of $\theta$, however there exists a necessary and sufficient
condition that guarantees the existence of an estimate achieving the
\crra\ bound, $\mathcal{I}_{\theta}(n)^{-1}$. More precisely, if we
can write (see, e.g., \citet[pp. 12]{stu04}):

\begin{equation}\label{decomp}
\frac{d \ln L(I_i,...,I_n;\theta)}{d \theta} =
A(\theta) \cdot \left( \hat{\theta} (I_1,...,I_n)) -
\theta \right)
\end{equation}

then, it is certain that $Var(\hat \theta
(I_1,...,I_n))=1/\mathcal{I}_{\theta}(n)$, as long as $A(\theta)$ is a
function {\it exclusively} of the parameter $\theta$ (and, in
particular it does not depend on the observables,
$I_i$)\footnote{Furthermore, in this scenario, it can be easily shown
  that $A(\theta)=\mathcal{I}_{\theta}(n)$.}. However, we must keep in
mind that, unless the condition in equation~(\ref{decomp}) is
satisfied, the minimum variance solution from equation~(\ref{minvar})
will have, in general, a variance strictly greater than
$\mathcal{I}_{\theta}(n)^{-1}$. This important result will be further
used in Section~\ref{sec_pract_estimators}.

\subsection{Position estimation: Astrometry} \label{subsec:post_est_astrometry}

The position estimation in astrometry is a slight variation of the
classical parameter estimation problem introduced in
Section~\ref{subsec:parameter_est}.  In this paper we focus on the 1-D
version, as it is more easily tractable from the numerical and
analytical point of view, while capturing all the key elements of the
problem. The extension to the 2-D case will be dealt with in a
forthcoming paper. However, as we shall see (see Sections~\ref{2d}
and~\ref{sec_pract_estimators}), it is expected that some
generalizations are possible from our 1-D results which are likely to
be approximately valid on the 2-D scenario in some cases.

In the 1-D case we have an array detector with $n$ pixels, in which we
measure the fluxes $\{I_i\}$ per pixel. We will assume that the total
{\it expected} (as opposed to measured) flux at each pixel on the
detector, given by a function $\lambda_i(x_c)$, will explicitly
depend on the location of the source on the array, denoted by $x_c$,
which is the parameter we want to determine (equivalent to the unknown
parameter $\theta$ of Section~\ref{subsec:parameter_est}). Of course,
this flux is not measured directly, because the actual observations,
$I_i$, are subject to noise. However, on photon counting devices, such
as CCDs, the measured flux follow a Poisson noise distribution, i.e.,
the $I_i$ are random variables driven by a Poisson distribution (this
determines the probability mass function $f_\theta$ introduced in
Section~\ref{sec:pre}), with expectation value given by
$\lambda_i(x_c)$. At this point, we note an important difference in
approach to that adopted in the work by \citet{leevan83}, in which
they have assumed a Gaussian noise per pixel, valid for an analog
detector, such as photographic plates. As we shall see, when the noise
is Gaussian, a maximum likelihood parameter determination reduces to
least squares, which is what they indeed adopted. Note however that
\cite{kin83} also adopted a least squares minimization, although for
CCDs the noise is not Gaussian, and therefore a maximum likelihood
solution {\it is not} equivalent to a least squares minimization (see
Section~\ref{sec_pract_estimators}).

In order to estimate the \crra\ bound in this situation, we first need
to verify that our likelihood function satisfies
equation~(\ref{constr}). The likelihood function of the 1-D array
observations will be given by:

\begin{equation}\label{like1}
L(I_1,...,I_n;x_c)=f_{\lambda_1(x_c)}(I_1) \cdot
f_{\lambda_2(x_c)}(I_2) \cdots f_{\lambda_n(x_c)}(I_n)
\end{equation}

where $f_{\lambda}(I)=\frac{e^{-\lambda}\cdot \lambda^I}{I!}$, since
the $I_i$ follow a Poisson mass function distribution with mean
$\lambda_i(x_c)$
\footnote{Note that this estimation setting is different from the
  classical setting of Section~\ref{subsec:parameter_est}, since we
  have random independent, although not identically distributed,
  samples. Nevertheless, it is simple to prove that
  equations~(\ref{varcr}) to (\ref{fisher}) still hold under this more
  general setting.}. Then, we see that:

\begin{eqnarray}
\frac{d \ln L(I_1,...,I_n;x_c) }{d x_c} & = & \frac{d}{d x_c}
\left( {\sum_{i=1}^n \left( I_i\cdot \ln \lambda_i(x_c) - \lambda_i(x_c) -
    \ln I_i! \right)  } \right) \label{derilike1} \\
& = & \sum_{i=1}^n I_i \cdot \frac{1}{\lambda_i(x_c)} \cdot \frac{d
  \lambda_i(x_c)}{d x_c} - \sum_{i=1}^n \frac{d \lambda_i(x_c)}{dx_c} \label{derilike}
\end{eqnarray}

and, we indeed verify that $\mathbb{E}_{I_1,...,I_n} \left( \frac{d
  \ln L(I_1,...,I_n;x_c)}{d x_c} \right) =0$ because
$\mathbb{E}({I_i})=\lambda_i(x_c)$. Hence, we can apply
equations~(\ref{varcr}) and (\ref{fisher}) to obtain the following
result:

\begin{equation} \label{var0}
Var (\hat{x}_c(I_1,...,I_n)) \geq \frac{1} { \mathcal{I}_{x_c}(n)} =
\frac{1} { \displaystyle \sum_{i=1}^n \frac{ \left( \frac{d
      \lambda_i(x_c)}{d x_c} \right)^2 } {\lambda_i(x_c)} }
\end{equation}

For completeness, the derivation of the Fisher information about
$x_c$, $\mathcal{I}_{x_c}(n)$, is presented in Appendix
\ref{app_fisher}.

In the following section, we provide a detailed analysis of this
expression and its practical implications on astrometry.

\section{The astrometric \crra\ minimum variance bound in 1-D}
\label{astrcrra}

In very general terms, the observed fluxes $\{I_i\}$ will have
contributions from the source itself, as well as from a
background. Correspondingly, the expected flux (in one pixel) from the
source (which explicitly depends on $x_c$) will be characterized by a
function $\tilde{F}_i(x_c)$, representing the flux (in photo-e$^-$ on
the detector) at pixel $i$, whereas the expected generic background
will be denoted by $\tilde{B}_i$, representing the total (integrated)
background (also in units of e$^-$) at pixel $i$, and which includes
contributions from the detector (read-out noise and dark-current, if
any), and the sky background. We will assume that $\tilde{B}_i$ does
not depend on $x_c$. The total expected flux will thus be given by
$\lambda_i(x_c) = \tilde{F}_i(x_c) + \tilde{B}_i$. If we replace this
expression for $\lambda_i(x_c)$ into equation~(\ref{var0}), we see
that:

\begin{equation}\label{var1}
Var(\hat{x}_c) \ge \sigma^2_{\mbox{\tiny CR}} = \frac{1}{\displaystyle
  \sum_{i=1}^{n} \frac{ \left( \frac{d \tilde{F}_i}{d x_c}(x_c)
    \right) ^2 } {\left( \tilde{F}_i(x_c) + \tilde{B}_{i} \right) }}
\end{equation}

At this point, it is convenient to define a dimensionless, normalized,
function $g_i(x_c)$ such that $\tilde{F}_i (x_c)= \tilde{F} \cdot
g_i(x_c)$, where $\tilde{F}$ is the total flux of the object (which is
invariant to the actual value of $x_c$). In this case, the RHS of
equation~(\ref{var1}) can be written as:

\begin{equation}
\sigma^2_{\mbox{\tiny CR}} = \frac{1}{\displaystyle \sum_{i=1}^{n}
  \frac{\left( \tilde{F} \, \frac{d g_i}{d x_c}(x_c) \right)
    ^2}{\left( \tilde{F} \, g_i(x_c) + \tilde{B}_{i} \right)
}} \label{var2}
\end{equation}

Note that this expression is similar to the \crra\ bound derived by
\citet{win86} (his equation~(35)).

The function $g_i(x_c)$ is determined by the {\it Point Spread
  Function} (PSF), which describes the distribution of (source) flux
on the detector or, equivalently, the image profile across pixels
(represented by the function $\Phi(x)$), integrated over pixel $i$ (of
width $\Delta x$) of the array, i.e.:

\begin{equation}
g_i(x_c) = \int_{x_i-\frac{\Delta
    x}{2}}^{x_i+\frac{\Delta x}{2}} \Phi(x) \, dx \label{resp}
\end{equation}

Note that the PSF function $\Phi(x)$ is also normalized, i.e.,

\begin{equation}
\int_{-\infty}^{+\infty} \Phi(x) \, dx = 1
\end{equation}

As long as the array length samples a significant fraction of the PSF,
we can indeed identify $\tilde{F}$ as the total flux of the star,
since:

\begin{equation}
\sum_{i=1}^n \tilde{F}_i = \tilde{F} \sum_{i=1}^n g_i(x_c) = \tilde{F}
\sum_{i=1}^n \int_{x_i-\frac{\Delta x}{2}}^{x_i+\frac{\Delta x}{2}}
\Phi(x) \, dx \approx \tilde{F} \int_{-\infty}^{+\infty} \Phi(x) \,
  dx = \tilde{F}
\end{equation}

For practical purposes, since the detector array length greatly
exceeds the PSF extent, this means that the source must not be too
close to the array boundaries for this equation to be valid.

\subsection{Interpretation of the structure of the \crra\ bound} \label{interp}

An equivalent to equation~(\ref{var1}) in 2-D has been presented, in a
slightly different manner, by \citet{leevan83} in the context of the
expected astrometric accuracy on photographic plates (their
equation~(9)). Indeed, it is easy to see that their $\Delta_{ij}^2
\equiv \tilde{F}_{ij}(x_c,y_c) + \tilde{B}_{ij} \equiv
\sigma_{{\tilde{F}}_{ij}}^2 + \sigma_{{\tilde{B}}_{ij}}^2$, where we
have assumed that both, the source and the background, follow Poisson
noise (in the case of CCD detectors this is valid when the fluxes are
measured in photo-e$^-$), and where $\sigma_{{\tilde{F}}_{ij}}^2$ and
$\sigma_{{\tilde{B}}_{ij}}^2$ are the variances in the source and the
background, respectively, at pixel $i,j$. We emphasize however that
their equation~(9) has been derived in the case of a (weighted) least
squares solution (their equation~(5)). This equation is valid for an
analog detector, such as photographic plates, where the photographic
densities in each pixel are assumed to follow Gaussian noise (see
equation~(4) in \citet{leevan83}). However, as it will be demonstrated
in Section~\ref{sec_pract_estimators}), in the case of digital
detectors, where the noise follows a Poisson distribution, no
parameter estimator can formally reach the \crra\ bound.

%

We note that, for Poisson noise, the standard deviation of the signal
is the square root of the signal and, thus, the variance is the signal
itself. Thus, the interpretation of the term $\tilde{F}_{i}(x_c) +
\tilde{B}_{i}$ in equation~(\ref{var1}) as the {\it variance} of the
counts (in e$^-$) is important, as indicated in what follows. It is
often more convenient for evaluation purposes to express $\tilde{F}$
and $\tilde{B}$ in terms of ``counts'' on the detector (from now on
referred to as ``ADUs'' - Analog to Digital Units), rather than in
e$^-$, by introducing the so-called (inverse-)gain of the detector $G$
in units of e$^-$/ADU (see, e.g., \citet{gill92}). In this case, the
source and background fluxes, $\tilde{F}$ and $\tilde{B}$, are given
by $G \cdot F$ and by $G \cdot B$ respectively, where $F$ and $B$ are
in units of ADU. On the other hand if $\sigma_{{\tilde{F}}_i}$ is the
rms deviation at pixel $i$ (in e$^-$) and $\sigma_{{F}_i}$ is the
equivalent quantity in units of ADUs, then it is true that
$\sigma_{{\tilde{F}}_i} = G \cdot \sigma_{{F}_i}$, and similarly for
the rms deviation on the background. By replacing these unit
conversions into either equation~(\ref{var1}) or (\ref{var2}) we see
that we can express those equations with the source and the background
measured in units of either e$^-$ or ADUs, in the sense that:

\begin{eqnarray}
\sigma^2_{\mbox{\tiny CR}} & = & \frac{1}{\displaystyle \sum_{i=1}^{n}
  \frac{ \left( \frac{d \tilde{F}_i}{d x_c}(x_c) \right) ^2 } {\left(
    \sigma^2_{{\tilde{F}}_i} + \sigma^2_{{\tilde{B}}_i} \right)
}} \label {var3a} \\ & = & \frac{1}{\displaystyle \sum_{i=1}^{n}
  \frac{ \left( \frac{d F_i}{d x_c}(x_c) \right) ^2 } {\left(
    \sigma^2_{{F}_i} + \sigma^2_{{B}_i} \right) }} \label{var3b}
\end{eqnarray}

To evaluate the \crra\ bound if we have empirical measurements of {\it
  fluxes and variances} on a detector (ADUs), it would obviously be
more convenient to use equation~(\ref{var3b}). However, when
performing numerical experiments for a given detector set-up (as done
in this paper, Section~\ref{numres}), where we only specify flux
levels for the source and the background, and where we {\it calculate}
the variances associated to them, we would instead need to use
equation~(\ref{var3a}) (see also equation~(\ref{exact})). This is
because we know that if the flux is in expressed e$^-$, then the
variances are equal to the flux (but this is not the case if the flux
is measured in ADUs). We further note that equations~(\ref{var3a}) or
(\ref{var3b}) suggest that the \crra\ bound represents a mean
``uncertainty over the derivate of the signal'', since:

\begin{equation}
\sigma^2_{\mbox{\tiny CR}} = \frac{1}{\displaystyle \sum_{i=1}^{n}
  \left( \frac{ \frac{d \tilde{F}_i}{d x_c}(x_c) }
       {\sigma_{\lambda_i}} \right)^2} = \left< \left(
\frac{\sigma_{\lambda}} { \left( \frac{d \tilde{F}}{d x_c} \right)}
\right) ^2 \right>
\end{equation}

where the $<>$ stands for a classical type of harmonic mean over the
pixels, and where $\sigma^2_{\lambda_i} = \sigma^2_{\tilde{F}_i} +
\sigma^2_{\tilde{B}_i}$. This provides an interesting connection with
what will be presented in Section~\ref{numres}: We note that if we
have a function of one variable, $y=f(x)$, then\footnote{Using a
  Taylor expansion around a point and assuming that $f$ is
  sufficiently regular or smooth around that point.} $\sigma_y^2
\simeq \left( \frac{df}{dx} \right)^2 \cdot \sigma_x^2$. If we have
$n$ independent measurements, indicated by ($x_i, y_i$), each with
uncertainty ($\sigma_{x_i},\sigma_{y_i}$) we know from the propagation
of errors in a least squares sense that the minimum variance for the
weighted mean $\bar{x}$ would be given by $\sigma_{\bar{x}}^2 =
\frac{1}{\displaystyle \sum_{i=1}^n
  \frac{1}{\sigma_{x_i}^2}}=\frac{1}{\displaystyle \sum_{i=1}^n \left(
  \frac{ \left( \frac{df_i}{dx} \right) }{\sigma_{y_i}} \right)^2 }$
(see, e.g., \citet{mey92}, Chapter~10), which is equivalent to the
\crra\ bound if we identify $f_i = \lambda_i$ and, since the errors
follow a Poisson distribution, then $\sigma^2_{y_i} = \lambda_i$. We
emphasize that this analogy is valid only in the 1-D case (otherwise
we would need to include the partial derivatives of $f$, and the
variances in all the parameters). Indeed, as we shall see in
Section~\ref{sec_pract_estimators}, the least squares does not reach,
in general, the \crra\ lower bound.

An interesting aspect of equations~(\ref{var1}) or (\ref{var2}) is
that the positional uncertainty is going to be dominated by the region
near the center of the PSF, where its derivative is steeper. Including
regions far from the central core of the PSF, where $\frac{d
  \tilde{F}_i}{d x_c}(x_c) \sim 0$, will not contribute to a decrease
in the uncertainty and, instead, will only deteriorate the overall
$S/N$ of the source by incrementally adding more noise than signal
(see the paragraph following equation~(\ref{sn3})).

\subsection{The \crra\ bound for a Gaussian source} \label{gausou}

It is evident that very little progress can be made in the estimation
of the \crra\ bound from equation~(\ref{var2}), unless we specify a
shape for the PSF. Various analytical forms have been proposed for the
PSF of a point source as imaged by ground-based \citep{kin71} and
space-based detectors \citep{kin83} (but, see also
\citet{ben88}).
Without loosing too much generality, in this study we will adopt a
Gaussian function which seems to be a good representation of the PSF,
at least from the stand-point of astrometric accuracy on ground-based
data (\citet{men10}), and which allows some simple analytical
manipulation (see Section~\ref{contcrra}). Under this assumption, we
would have:

\begin{equation}
\Phi(x) = \frac{1}{\sqrt{2 \pi} \, \sigma} \, e^{-\frac{(x-x_c)^2}{2
    \, \sigma^2}} \; \; \mbox{arcsec}^{-1}\label{psf}
\end{equation}

where we adopt to measure $x$, $x_c$ and $\sigma$ in units of arcsec.

In this case, it is easy to show that the derivative of $g_i$ in
equation~(\ref{resp}) can be written in a closed-form, as follows:

\begin{equation}
\frac{d g_i}{d x_c}(x_c) = \frac{1}{\sqrt{2 \pi} \, \sigma} \left(
e^{-\gamma(x^-_i)} - e^{-\gamma(x^+_i)} \right) \; \;
\mbox{arcsec}^{-1} \label{deriv}
\end{equation} 

where:

\begin{equation}
\gamma(x) = \frac{(x-x_c)^2}{2 \, \sigma^2} \label{gamma}
\end{equation}

with $x^{\_}_i = x_i - \frac{\Delta x}{2}$ and $x^+_i= x_i +
\frac{\Delta x}{2}$.

Combining the results of equations~(\ref{resp}), (\ref{psf}),
(\ref{deriv}) and (\ref{gamma}) in (\ref{var2}) and converting to ADUs,
we finally arrive at the following exact expression for the
\crra\ lower-bound in 1-D for a Gaussian PSF:

\begin{equation}
\sigma^2_{\mbox{\tiny CR}} = 2 \pi \sigma^2 \cdot \frac{B}{G \, F^2}
\cdot \frac{1}{\displaystyle \sum_{i=1}^{n} \frac{\left(
    e^{-\gamma(x^-_i)} - e^{-\gamma(x^+_i)} \right) ^2}{\left( 1 +
    \frac{1}{\sqrt{2 \pi} \, \sigma}\frac{F}{B} \displaystyle
    \int_{x_i-\frac{\Delta x}{2}}^{x_i+\frac{\Delta x}{2}}
    e^{-\gamma(x)} \, dx \right) } } \label{exact}
\end{equation}

where we have assumed, for simplicity, that the background is uniform
(and equal to $B$) under the PSF of the object. As noted by
\citet{win86}, this expression makes it explicit that the \crra\ bound
depends on $F$ and $B$ separately, and not just on the ratio $F/B$. If
we adopt $\sigma$ and $\Delta x$ in unit of arcsec in the sky, then
the square-root of equation (\ref{exact}) gives us the \crra\ bound in
units of arcsec directly.

In Figure~\ref{figcr1} we show the results of evaluating equation
(\ref{exact}) under the experimental setting proposed by
\citet{win86}, i.e., assuming a fixed set of $F$ and $B$ values (and
therefore a constant ratio $F/B$), for different values of the
detector pixel size $\Delta x$. In this figure we introduce the
``Full-Width at Half-Maximum'' ($FWHM$), usually termed as ``image
quality'' at astronomical observing sites, which is related to the
Gaussian $\sigma$ through $FWHM = 2 \sqrt{2 \ln 2} \,\, \sigma$.
Figure~\ref{figcr1} is equivalent to Figure~1 in \citet{win86}.

We note that, for a given continuous pixel $x$-coordinate on the
array, the corresponding (integer) pixel ID, $i$, on the array, is
given by:

\begin{equation}
i = INT \, (x+0.5)
\end{equation}

where the function $INT$ represents the integer part of the
argument. In this paper we adopt that pixel ID $i=1$ has pixel
coordinates $0.5 \le x < 1.5$, pixel ID $i=2$ has pixel coordinates
$1.5 \le x < 2.5$, and so on, following the convention of the IRAF
package\footnote{IRAF is distributed by the National Optical Astronomy
  Observatories, which are operated by the Association of Universities
  for Research in Astronomy, Inc., under cooperative agreement with
  the National Science Foundation, see http://iraf.noao.edu/.}. In
this scheme, each pixel has width 1.0 in pixel $x$-coordinates,
centered at $x= FLOAT\,(i)$, with upper/lower pixel boundaries given
by $x_{\pm} = FLOAT\,(i) \pm 0.5$ (where the $FLOAT$ function converts
an integer into a real number). The relationship between pixel
$x$-coordinates and ``physical'' coordinates (projected onto the sky,
as measured from the origin of the array), is given by $px = [(x-0.5)
  \cdot \Delta x]$~arcsec.

\subsection{The \crra\ bound and dithering in undersampled images} \label{dith}

An interesting feature is the effect on the predicted \crra\ bound of
pixel de-centering of the source. Figure~\ref{figcr1} shows the effect
of pixel de-centering for a $FWHM=0.5$~arcsec, as a function of
$\Delta x$. The fact that the \crra\ bound {\it depends} on the
location of the source itself was already pointed-out by
\citet{win86}, but this is evident from equations~(\ref{var1}),
(\ref{var2}) or (\ref{exact}), all of which explicitly depend on
$x_c$. Of course, the effect is symmetrical with respect to the pixel
center, so the impact of a de-centering of, e.g., +0.125~pix with
respect to the pixel center is exactly the same as that of a
de-centering of -0.125~pix, and (as long as the array properly samples
the source), i.e. the effect is periodic (such that the \crra\ bound
is the same if the source is placed at $x \pm FLOAT\,(n)$, where $n$
is an arbitrary integer).

The important role of image de-centering in the case of undersampoled
images (such as those of $HST$, see below), has been demonstrated by
\citet{andkin00}. They refer to the ``pixel$-$phase error'' as the
systematic error in the derived center if a centering algorithm is
used that does not properly account for the distribution of signal
among pixels when the source is not centered in a pixel. On the other
hand, the \crra\ bound described here represents instead the
precision, or random error, statistically attainable with an ideal,
unbiased, centering algorithm.

The dotted line on Figure~\ref{figcr1} is centered on a pixel, near
the center of the array, the double-dotted-dashed line is for an
offset of 0.125~pix, and the dot-dashed line is for a pixel
de-centering of 0.25~pix. As it can be seen, the loss of astrometric
accuracy as the pixel size increases, is less severe when the target
is {\it not} at the center of a given pixel, but, rather, when it is
offset from it. This is actually an intuitive result: For a given
pixel size, when the source is not centered on a pixel, its flux is
spread among more neighboring pixels, and therefore the source can be
located more precisely. This result also implies that, for
under-sampled systems, it is a good practice to ``dither'' the source
a bit (even a fraction of a pixel) so that, in the end, the average
\crra\ bound is better than that if the source were located at the
center of a pixel, a well known technique applied, e.g., to $HST$
images (\citet{andkin00}, \citet{fruhoo02}). To quantify the effect of
dithering, in Table~\ref{tabdit} we show the \crra\ bound for a source
with a $FWHM=0.5$~arcsec observed through detectors of pixel size
$\Delta x=0.5$, 0.7 and 0.9~arcsec, and when the source is centered
and slightly de-centered by the amounts indicated in the table. As it
can be seen from this table, for a pixel size that matches the $FWHM$,
the change on the \crra\ bound is small as a function of pixel
offset. In the case of a 0.7~arcsec pixel size, a dither pattern
including offsets of 0.125 and 0.25~pix, plus a central pointing,
yields an average \crra\ bound of $\sim$6.0~mas, which represents an
almost 18\% improvement over the (single) centered \crra\ bound. For a
0.9~arcsec detector the effect is even more dramatic, yielding an
almost 40\% improvement.

Figure~\ref{figcr1} also shows that, while the dithering technique
offers a better asymptotic trend among all the values at large $\Delta
x$ (i.e., in the low-resolution regime), eventually, when $\Delta x$
becomes too large, the astrometric accuracy deteriorates regardless of
the relative position of the source with respect to the center of a
pixel (even with dithering), as expected.

\section{Astronomical application of the \crra\ bound} \label{astrappl}

We must note that $F$ in equation~(\ref{exact}) is the total source
flux, which is independent of the pixel size on the detector, whereas
$B$ is, instead, the background level {\it in one pixel}. Therefore,
as $\Delta x$ becomes smaller and smaller, the {\it total}
contribution from the background under the PSF of the source increases
steadily (because $B$ is fixed, and the number of pixels under the PSF
increases), and the positional precision deteriorates (for a specific
connection with the $S/N$ ratio of the source, see
equation~(\ref{sn3}) and the comments that follow that equation). On
the other extreme, as $\Delta x$ increases, we loose resolution and
the positional precision also deteriorates. We get a ``valley'' which
determines an optimum region $\Delta x$ for a given set of $F$, $B$
(and $\sigma$). While this setting is interesting in certain
applications where the value of $B$ is independent of the pixel size
(e.g., military or day-time applications where the readout of the
array is very fast, and the background is dominated by the electronics
of the device, or when we are dominated by dark-current, see
equation~(\ref{back})), the situation in astronomical applications is
quite different: In this case, the long readout times imply that, in
most cases, the background is dominated by diffuse light coming from
the sky, and not from the detector, and in this case the background in
a given pixel {\it is not} independent of the pixel size, as assumed
in the analysis by \citet{win86}. The setting for evaluating
equation~(\ref{exact}) must then be adapted to our case of interest,
this is done in the next section.

As indicated in the previous paragraph, the correct expression for the
(constant as a function of $x$, see equation~(\ref{exact})) background
$B$ contained in one pixel, is given in this case by:

\begin{equation}
B = f_s \, \Delta x + \frac{D + RON^2}{G} \; \;
\mbox{[ADU]} \label{back}
\end{equation}

where $f_s$ is the sky background (in units of ADUs/arcsec), while $D$
and $RON$ are the dark-current and read-out noise of the detector
\citep[pp. 222]{how13}\footnote{See also, e.g.,
  http://www.ucolick.org/$\sim$bolte/AY257/s\_n.pdf, last accessed on
  December 2012.}, per pixel, in units of e$^-$. In the paper by
\citet{win86} it was assumed that $f_s \sim 0$ (true for very short
exposure times), and we can see that, indeed, in this case, the
background is independent of $\Delta x$. In what follow we will
neglect the contribution from dark-current, which in current CCD
detectors is negligible.

With this new prescription for $B$, and in order to evaluate the RHS
of expression~(\ref{exact}) for some astronomically interesting
situations, it is is also worth to develop some easily {\it
  measurable} form of ``signal'' and ``noise'' for our Gaussian
source, as observed through our CCD detector. In this case one could
define the signal $S$, as:

\begin{equation}
S = G \cdot F \cdot \int_{x_l}^{x_u} \Phi(x) \, dx
\;\; [e^-] \label{signal1}
\end{equation}

where $x_l$ and $x_u$ are suitably chosen (but arbitrary) apertures
that include an appreciable fraction of the total flux of the star (we
can not actually measure from $-\infty$ to $+\infty$ with a real
detector, nor we want to do that since, in this formulation, the
background will add-up to infinity over that aperture as well, see
equations~(\ref{noise}) and (\ref{npix})). For the case of the
Gaussian function adopted here, equation~(\ref{psf}), it makes sense
to perform an integration of the PSF that is symmetrical with respect
to the center of the source, centered at $x_c$, in which case the
signal can be written as:

\begin{equation}
S = G \cdot F \cdot P(u_+) \;\; [e^-] \label{signal2}
\end{equation}

where $P(u_+)$ is the probability integral\footnote{The probability
  integral is defined as $P(u)=\frac{2}{\sqrt{\pi}} \int_{0}^{u}
  e^{-v^2} \, dv$} evaluated at $u_+ = \left( x_u - x_c
\right)/\sqrt{2} \, \sigma$, and where $x_u - x_c = x_c - x_l$.

The total noise, $N$, has contributions from the read-out-noise of the
detector, the noise from the sky, and the noise from the source
itself, all of which are assumed to follow Poisson statistics (in
e$^-$), such that (see, e.g., \citet{gill92}):

\begin{equation}
N = \sqrt{ S + N_{\mbox{pix}} \left( {G f_s \, \Delta x + RON^2
  } \right) } \;\; [e^-] \label{noise}
\end{equation}

where $N_{\mbox{pix}}$ is the number of pixels under the same region
in which the signal $S$ was sampled, i.e., in the interval
$[x_l,x_u]$, which is given by:

\begin{equation}
N_{\mbox{pix}} = \frac{x_u - x_l}{\Delta x} =
\frac{2 \, \sqrt{2} \, \sigma \, u_+}{\Delta x} = \frac{u_+}{\sqrt{\ln
    2}} \cdot \frac{FWHM}{\Delta x} \label{npix}
\end{equation}

Combining equations~(\ref{signal2}), (\ref{noise}) and (\ref{npix}) we see that:

\begin{eqnarray}
\frac{S}{N} (u_+) & = & \frac{ P(u_+) \cdot F}{\sqrt{\frac{P(u_+)
      \cdot F}{G} + \frac{2 \, \sqrt{2} \, u_+}{G}
    \frac{\sigma}{\Delta x} \left( f_s \Delta x + \frac{RON^2}{G}
    \right) }} \nonumber \\
            & = & \frac{ P(u_+) \cdot F}{\sqrt{\frac{P(u_+) \cdot
      F}{G} + \frac{u_+}{\sqrt{\ln 2} \, G} \frac{FWHM}{\Delta x}
    \left( f_s \Delta x + \frac{RON^2}{G} \right)
}} \label{sn3}
\end{eqnarray}

For example, for $P=0.9$ (aperture containing 90\% of the total flux),
we have $u_+ \sim 1.164$ (the true ``physical'' aperture on the
detector would be $\frac{1.164}{\sqrt{\ln 2}} \times FWHM \approx 1.40
\times FWHM$), whereas for $P=0.99$ then $u_+ \sim 1.822$ (or
$\frac{1.822}{\sqrt{\ln 2}} \times FWHM \approx 2.19 \times
FWHM$). Because $u_+$ increases faster than $ P(u_+)$ (which is bound
to a maximum value of 1.0), we see that, for a given source,
background, and detector, the $S/N$ computed from equation~(\ref{sn3})
decreases as $u_+$ increases beyond the main core of the PSF. For
example, for $\Delta x =0.2$~arcsec, $G=2$~e$^-$/ADU, $RON=5$~e$^-$,
$f_s = 2\,000$~ADU/arcsec, $F=5\,000$~ADU, and $FWHM=1.0$~arcsec, then
for $P=0.9$, $S/N \sim 74$, whereas for $P=0.999$ (for which $u_+ \sim
2.33$), $S/N \sim 68$. As was explained before, we note that
equation~(\ref{exact}) does not depend directly on the $S/N$.

Equation~(\ref{sn3}) is interesting since it explicitly shows that, as
$\Delta x$ becomes smaller and smaller, the $RON$ term starts to
dominate over the sky background in its contribution to the total
noise,
the impact of which, on the \crra\ bound, has already been mentioned
in Subsection~\ref{gausou}. However, when $\Delta x$ increases, the
sky background becomes the dominant source of background noise, and
the total noise becomes {\it independent} of the array pixel
size. Also, this equation clearly shows the classical result that, as
an image becomes more spread (larger $FWHM$, or worse image quality)
the $S/N$ deteriorates, for a fixed total flux $F$, because of the
larger contribution from the sky and the (larger number of) pixels
underneath the aperture: As we shall see, the $FWHM$ has a very
relevant impact on the \crra\ bound (see, e.g.,
equation~(\ref{weakstrong})).

Figure~\ref{figcr2} shows the result of evaluating
equation~(\ref{exact}) under the assumption of a background given by
equation~(\ref{back}) for a set of representative values. An
interesting point here is that, at very small values of $\Delta x$ we
still see the ``upturn'' in the \crra\ lower bound seen in
Figure~\ref{figcr1}, but it has a much smaller effect. Of course, the
reason for this upturn is the prevalence of the $RON$ over the sky
background indicated in the previous paragraph, when $\Delta x$
becomes extremely small. As we shall see
(equation~(\ref{weakstrong})), the \crra\ bound goes as $\Delta x
^{-1}$ for small $S/N$ and small pixels, a feature clearly seen in
Figure~\ref{figcr2}. Otherwise we see a broad region that exhibits a
rather smooth and steady decrease in positional precision when $\Delta
x$ becomes larger and larger, and a rather steep increase when $\Delta
x$ increases beyond the $FWHM$. The overall effects of pixel
de-centering are qualitatively similar to those already presented in
Figure~\ref{figcr1}, and are thus not repeated in this figure. For
very large $S/N$, equation~(\ref{weakstrong}) predicts that the
\crra\ bound becomes rather insensitive to $\Delta x$, which also
coincides with the behavior in Figure~\ref{figcr2}.

An interesting prediction of equation~(\ref{exact}) is that
high-resolution imaging in low-background, even for under-sampled
images (e.g., HST), is better than imaging with larger aperture
ground-based telescopes, not under-sampled, due to the worse $FWHM$
and higher-background of the latter, a well-known fact by people doing
astrometry with HST (provided, of course, that systematic effects are
well understood, e.g., a particularly challenging situation with HST
data is the account of time-dependent charge-transfer efficiency
corrections, for details see, e.g., \citet{bri05}, especially their
Figure~4, or \citet{bri06}, especially their Figure~10). For example,
for the same detector parameters as those adopted in
Figure~\ref{figcr2}, and $F=10\,000$ (which for a Gaussian PSF leads
to maximum flux in the central pixel of $\sim 1\,700$~ADU (see
Section~\ref{contcrra} and equation~(\ref{fmaxapp})),
$f_s=3\,000$~ADU/pix, and $FWHM=0.45$~arcsec, the \crra\ bound is
$\sim$1.7~mas (with $\Delta x = 0.08$~arcsec). These (source \& image)
values are similar to those of the QSOs used in the astrometric study
by \citet{men10} (see their Table~1) and \citet{men11}, which
demonstrated a single-measurement astrometric precision of 1.5~mas
(see Section~3.2 in \citet{men10}) with the NTT (3.5m aperture)
telescope and SUSI2 imager. On the other hand, for HST with
$f_s=30$~ADU/pix and $FWHM=0.15$~arcsec, then the \crra\ bound is
$\sim$0.2~mas (in this case $\Delta x = 0.1$~arcsec), whereas
\citet{pia02} reported a single-measurement precision of 0.25~mas (in
our calculation of the \crra\ bound for HST we have approximately
taken into account the aperture difference between the NTT and HST,
and the different exposure times for the same QSOs adopted in these
two studies, from Table~1 in \citet{pia02}).

\subsection{The \crra\ bound in the small pixel (high resolution) approximation} \label{contcrra}

Under certain circumstances, the summation in the denominator of the
RHS of equation~(\ref{var1}) can be approximated into an integral,
which allows us to explore the behavior of the \crra\ bound in a more
explicit manner. Indeed, we see from equations~(\ref{resp}) and
(\ref{psf}) that the application of the mean-value theorem when
$\Delta x / \sigma \ll 1$ implies that $\tilde{F}_i \equiv \tilde{F}
\cdot g_i(x_c) \approx \tilde{F} \cdot \Phi(x_i) \cdot \Delta
x \label{respapp}$. In this case, for a Gaussian PSF, it is easy to
show that:

\begin{equation} \label{derivap}
\frac{d \tilde{F}_i}{d x_c}(x_c) \equiv \tilde{F}
\frac{d g_i}{d x_c}(x_c) = \frac{( x_i - x_c
  )}{\sigma^2} \cdot \tilde{F}_i
\end{equation}

Replacing the RHS of equation~(\ref{derivap}) into the RHS of
equation~(\ref{var1}) we have:

\begin{equation}
\sigma^2_{\mbox{\tiny CR}} = \sigma^4 \cdot \frac{1}{ \sum_{i=1}^{n} \left(
  x_i - x_c \right) ^2 \cdot \frac{\tilde{F}_i^2} {\left( \tilde{F}_i
    + \tilde{B}_{i} \right) } } \label{varapp1}
\end{equation}

Let us have a closer look at the (dimensionless) $\tilde{F}_i$ and
$\tilde{F}_i^2$ terms in the RHS of equation~(\ref{varapp1}). For our
Gaussian function we will have:

\begin{equation}
\tilde{F}_i \equiv \tilde{F} \int_{x_i-\frac{\Delta
    x}{2}}^{x_i+\frac{\Delta x}{2}} \Phi(x) \, dx =
\frac{\tilde{F}}{{\sqrt{2\pi} \sigma}} \int_{x_i-\frac{\Delta
    x}{2}}^{x_i+\frac{\Delta x}{2}} e^{-\frac{(x-x_c)^2}{2 \,
    \sigma^2}} \, dx \label{flux}
\end{equation}

which, in the small pixel size approximation becomes:


\begin{eqnarray}
\tilde{F}_i & \approx & \frac{\tilde{F}}{{\sqrt{2\pi} \sigma}} \cdot
e^{-\frac{(x_i-x_c)^2}{2 \, \sigma^2}} \cdot \Delta x \label{fluxapp1}
\\ & = & \tilde{F}_{\mbox{max}} \cdot e^{-\frac{(x_i-x_c)^2}{2 \,
    \sigma^2}} \label{fluxapp2}
\end{eqnarray}

where $\tilde{F}_{\mbox{max}}$ is the (dimensionless) maximum flux,
which will occur at a certain pixel $j$ that satisfies the condition
$x_j - \Delta x/2 \le x_c < x_j + \Delta x/2$.

Equation~(\ref{fluxapp1}) prompts us to define a {\it new function},
which describes the distribution across pixels, in the small-pixel
approximation, of the {\it square of the flux} that appears in
equation~(\ref{varapp1}), as:

\begin{equation}
\tilde{F}_i^2 = \hat{F}^2 \cdot e^{-\frac{(x_i-x_c)^2}{\sigma^2}} \cdot
\Delta x \label{fluxsqapp}
\end{equation}

where $\hat{F}^2$ is a proper normalization factor (in units of
arcsec$^{-1}$, see below).

Combining equations~(\ref{fluxapp2}) and~(\ref{fluxsqapp}) in
equation~(\ref{varapp1}), and considering that $\Delta x$ is very
small, we have:

\begin{equation}
\sigma^2_{\mbox{\tiny CR}} = \frac{\sigma^4}{\hat{F}^2} \cdot
\lim_{\Delta x \rightarrow 0}
\frac{1}{\displaystyle \sum_{i=1}^{n} \frac{\left( x_i - x_c \right) ^2 \cdot
    e^{-\frac{(x_i-x_c)^2}{\sigma^2}}}{\left( \tilde{F}_i +
    \tilde{B}_{i} \right) } \cdot \Delta x } = 
\frac{\sigma^4}{\hat{F}^2} \cdot
\frac{1}{\displaystyle \int_{-\infty}^{+\infty} \frac{\left( x - x_c \right) ^2
    \cdot e^{-\frac{(x-x_c)^2}{\sigma^2}} }{\left( \tilde{F}(x) +
    \tilde{B}(x) \right) } \, dx } \label{varapp2}
\end{equation}

Taking into account the definition of $\tilde{F}_{\mbox{max}}$ on
equation~(\ref{fluxapp2}), from equation~(\ref{fluxsqapp}) we will
have that $\tilde{F}_{\mbox{max}}^2 \equiv \tilde{F}_j^2 \approx
\hat{F}^2 \cdot \Delta x$, where we have used the fact that, since
$\Delta x$ is very small, then $x_j \approx x_c$. Replacing this
approximation and equation~(\ref{fluxapp2}) into
equation~(\ref{varapp2}) we get for the \crra\ bound in the
small-pixel approximation:

\begin{equation}
\sigma^2_{\mbox{\tiny CR}} = \frac{\sigma^4}{\tilde{F}_{\mbox{max}}^2} \cdot
\frac{\Delta x}{\displaystyle \int_{-\infty}^{+\infty} \frac{\left( x - x_c \right)
    ^2 \, e^{-\frac{(x-x_c)^2}{\sigma^2}}
  }{\left(\tilde{F}_{\mbox{max}} \cdot e^{-\frac{(x-x_c)^2}{2 \,
        \sigma^2}}+ \tilde{B}(x) \right) } \, dx } \equiv
\frac{\sigma^4}{\tilde{F}_{\mbox{max}}^2} \cdot \frac{\Delta
  x}{I} \label{varapp3}
\end{equation}

where we note that (the trivial definition of) integral $I$ has units
of arcsec$^3$.

This formula is only valid under the assumption of small pixels, so,
in a general application, equation~(\ref{var1}) should be used
instead, or equation~(\ref{exact}) for a Gaussian PSF. However, the
truly interesting aspect of this equation is that it can be
explicitly evaluated in two extreme cases: When the detection is
dominated by the source, and when it is dominated by the
background. This is done in the next subsections.

\subsubsection{Weak source}\label{weasou}

When the background dominates, then $\tilde{F}_{\mbox{max}} \ll
\tilde{B}(x)$. In this case we would have:

\begin{equation}
I= \int_{-\infty}^{+\infty} \frac{\left( x - x_c \right) ^2 \,
  e^{-\frac{(x-x_c)^2}{\sigma^2}} }{\left(\tilde{F}_{\mbox{max}} \cdot
  e^{-\frac{(x-x_c)^2}{2 \, \sigma^2}}+ \tilde{B}(x) \right) } \, dx
\approx \int_{-\infty}^{+\infty} \frac{\left( x - x_c \right) ^2 \,
  e^{-\frac{(x-x_c)^2}{\sigma^2}} }{\tilde{B}(x)} \, dx
\end{equation}

If we assume an approximately constant background under the PSF of the
target then:

\begin{equation}
I = \frac{1}{\tilde{B}} \int_{-\infty}^{+\infty} \left( x - x_c \right) ^2 \,
e^{-\frac{(x-x_c)^2}{\sigma^2}} \, dx = \frac{\sqrt{\pi}}{2} \cdot
\frac{\sigma^3}{\tilde{B}}
\end{equation}

Therefore, in this case, replacing $I$ into equation~(\ref{varapp3})
and re-arranging terms, the \crra\ bound becomes:

\begin{eqnarray}
\sigma^2_{\mbox{\tiny CR}} & = &\frac{2}{\sqrt{\pi}} \cdot
\frac{\tilde{B}}{\tilde{F}_{\mbox{max}}^2} \cdot \Delta x \cdot \sigma \nonumber \\
& = & \frac{1}{\sqrt{2 \pi \ln 2}} \cdot
  \frac{B}{G \, F_{\mbox{max}}^2} \cdot \Delta x \cdot FWHM  \label{weak}
\end{eqnarray}

The (pedagogical) use of this equation is that it allows us to draw
some basic conclusions regarding the expected positional accuracy in
this regime. First, because $\tilde{F}_{\mbox{max}} \ll \tilde{B}(x)$
we predict, in general, a rather large positional uncertainty, as
expected, due to the low $S/N$ of the source. Furthermore, the
accuracy will improve proportionally to $F_{\mbox{max}}^{-1}$, in
agreement with equation (10) of \citet{leevan83} (compare also the
first line of equation~(\ref{weak}) with equation~(7) in
\citet{auevan78}). Furthermore, as intuitively expected, the accuracy
deteriorates for a larger background, coarser pixel size, or for
lower-quality images (or sites), but relatively slowly: Only as the
square root of these parameters.

\subsubsection{Strong source}\label{strsou}

In this case, the signal from the source dominates over the
background, i.e., $\tilde{F}_{\mbox{max}}\gg \tilde{B}(x)$, and the
approximation for $I$ becomes:

\begin{equation}
I= \int_{-\infty}^{+\infty} \frac{\left( x - x_c \right) ^2 \,
  e^{-\frac{(x-x_c)^2}{\sigma^2}} }{\left(\tilde{F}_{\mbox{max}} \cdot
  e^{-\frac{(x-x_c)^2}{2 \, \sigma^2}}+ \tilde{B}(x) \right) } \, dx
\approx \frac{1}{\tilde {F}_{\mbox{max}}} \int_{-\infty}^{+\infty}
\left( x - x_c \right) ^2 \, e^{-\frac{(x-x_c)^2}{2 \, \sigma^2}} \,
dx
\end{equation}

Evaluating the definite integral we have:

\begin{equation}
I= \sqrt{2 \pi} \cdot \frac{\sigma^3}{\tilde {F}_{\mbox{max}}}
\end{equation}

Replacing this value for $I$ into equation~(\ref{varapp3}) we end up
with:

\begin{eqnarray}
\sigma^2_{\mbox{\tiny CR}} & = & \frac{1}{\sqrt{2 \pi}} \cdot
\frac{1}{\tilde{F}_{\mbox{max}}} \cdot \Delta x \cdot \sigma \nonumber \\
 & = & \frac{1}{4 \sqrt{\pi \ln 2}} \cdot
\frac{1}{G \, F_{\mbox{max}}} \cdot \Delta x \cdot FWHM \label{strong}
\end{eqnarray}

We see that, in this regime, the ultimate positional accuracy is
proportional to $F_{\mbox{max}}^{-1/2}$, similarly to what was found
by \citet{leevan83} (see their equation~(13)). Again, the accuracy
deteriorates slowly for coarser pixel size and for lesser-quality
images, but in this case the background level, formally, plays no role
in the expected accuracy.

\subsubsection{Limiting cases as a function of total flux}\label{weastr}

Equations~(\ref{weak}) and (\ref{strong}) are a bit misleading because, by
definition, $F_{\mbox{max}}$ depends itself on the adopted value for
$\Delta x$ and, therefore, needs to be evaluated in each particular
case. However, in the small pixel approximation, one can find an
approximate relationship between the total flux $F$ (which is
independent of $\Delta x$) and $F_{\mbox{max}}$. Indeed, assuming, as
done before, that, $x_j \approx x_c$, then:

\begin{eqnarray}
F_{\mbox{max}} \equiv F_j & = & \frac{F}{{\sqrt{2\pi}
    \sigma}} \int_{x_j-\frac{\Delta x}{2}}^{x_j+\frac{\Delta x}{2}}
e^{-\frac{(x-x_c)^2}{2 \, \sigma^2}} \, dx \label{fmax} \\
& \approx & \frac{F}{\sqrt{2\pi}} \cdot \frac{\Delta x}{\sigma} \label{fmaxapp}
\end{eqnarray}

Replacing equation~(\ref{fmaxapp}) into equations~(\ref{weak})
and~(\ref{strong}) we have:

\large
\begin{equation}
\sigma^2_{\mbox{\tiny CR}} \approx \left\{
\begin{array}{cc}
\frac{\sqrt{\pi}}{2 \, (2 \, \ln 2)^{3/2}} \cdot \frac{B}{G \, F^2} \cdot
\frac{FWHM^3}{\Delta x} & \mbox{if $F \ll B$} \\
\frac{1}{8 \, \ln 2} \cdot \frac{1}{G \, F} \cdot FWHM^2 & \mbox{if $F
  \gg B$}
\end{array}
\right. \label{weakstrong}
\end{equation}
\normalsize

Interestingly, we now see that the positional accuracy should
deteriorate linearly with the $FWHM$ of the image for strong sources,
but it will not otherwise depend on the array pixel size. However, for
weak sources, the dependence on the $FWHM$ is not only steeper but,
also, the finer the pixel size, the larger the expected positional
uncertainty. This latter result should also be intuitive since, when
we are dominated by the background, the more pixels we have under the
PSF, the larger the contribution of the background will be, to the
point of significantly perturbing the final positional
accuracy. Equivalently, note that the term $B/\Delta x$ is the
background per unit area (see equation~(\ref{back})): The larger this
value becomes, we indeed expect a larger positional uncertainty, as
shown by the first line on equation~(\ref{weakstrong}).  The behavior
depicted by equation~(\ref{weakstrong}) is similar to that found by
\citet{kin83}: For sources where the background dominates, his
equation~(23) predicts that the positional uncertainty goes as
$\sqrt{B}/F$, whereas when the background is negligible, his
equation~(24) shows that the positional uncertainty increases as
$1/\sqrt{F}$. Similarly, \citet{lin78} (see also
\citet[pp. 127]{vafo13}) finds that for a seeing-limited image with no
background, the limiting astrometric precision goes as $FWHM/(S/N)$,
which is equivalent to equation~(\ref{weakstrong}) for the case $F >>
B$, when $S/N \sim \sqrt{F}$, see equation~(\ref{sn3})). All these
predictions coincide with those from our
equation~(\ref{weakstrong}). Finally, and as already noted in
Section~\ref{astrappl}, the general trends seen in Figure~\ref{figcr2}
are well explained by equation~(\ref{weakstrong}).

\subsubsection{Range of use of the high resolution \crra\ bound} \label{range}

Besides their qualitative usefulness, what is the range of
applicability of equations~(\ref{weak}) and (\ref{strong}), or
(\ref{weakstrong})? For illustration purposes, in Table~\ref{tabcomp1}
we compare the values predicted by these equations, with the ``exact''
prediction from equation~(\ref{exact}), for some representative values
of the parameters. From this Table we can see that, even for
relatively large pixels, in comparison with the $FWHM$, these
equations predict very reasonable values in comparison with the
``exact'' value, as long as we respect the conditions for $F/B$ under
which equations~(\ref{weak}) and (\ref{strong}) were derived. The
second and third lines show that, indeed, as predicted by
equation~(\ref{weakstrong}), the \crra\ bound does not depend on
$B$. The third and fourth lines show that in the high-$S/N$ regime the
\crra\ bound goes linearly with the $FWHM$, while a comparison of the
first and fifth line demonstrate that in the low-$S/N$ case the
\crra\ bound varies as the ratio $FWHM^2/\Delta x$, as shown by
equation~(\ref{weakstrong}). Finally, in the last two lines of the
table, we show a case when equation~(\ref{weakstrong}) fails
miserably, i.e., for an intermediate $S/N$ value.

\subsubsection{A simplified extension to the 2-D Case} \label{2d}

It can be intuitively argued from either equations~(\ref{weak})
and~(\ref{strong}), or equation~(\ref{weakstrong}), that in the 2-D
case the numerical factors in front of these equations will be
somewhat altered, but their basic dependence on $F$ and $B$ should be
basically maintained, as already indicated by a comparison of our
results to the 2-D results by \citet{auevan78}, \citet{leevan83} and
\citet{kin83} noted in Sections~\ref{weasou}, \ref{strsou} and
\ref{weastr}. For example, if the source is relatively symmetrical,
such that we can replace an x-y integration (similar to that of
equation~(\ref{varapp3}) but in the two array dimensions x, y) by a
polar-radial integration, we will end up with a $\Delta r$ instead of
a $\Delta x$, or even something like $\sqrt{\Delta x \cdot \Delta y}$
for a non-square pixel array. Therefore, even though our present
results are based purely on a 1-D array, they have a very interesting
predictive power, a point to which we will return in
Section~\ref{numres} (see also Table~\ref{tabsto}).

\section{Comparing the \crra\ bound with the performance of practical estimators} \label{sec_pract_estimators}

Let us remember that the necessary and sufficient condition for the
existence of an estimator that achieves the \crra\ bound is that the
likelihood function can be decomposed as in
equation~(\ref{decomp}). Unfortunately, equation~(\ref{derilike}) does
not offer, in general, the normal form of equation~(\ref{decomp}), and
consequently no estimator achieves the \crra\ lower bound.

To further explore this, it is illustrative to consider the
high-resolution regime of Section~\ref{contcrra}, in which case
equation~(\ref{derivap}) holds. Replacing this into
equation~(\ref{derilike}), and re-arranging some terms, we have:

\begin{equation}\label{prachigh}
\frac{d\ln L(I_1,...,I_n;x_c)}{d x_c} = \sum_{i=1}^{n}
\frac{\tilde{F_i}}{\sigma^2} \cdot \left( \frac{I_i}{\tilde{F}_i +
  \tilde{B}_i}-1 \right) \cdot (x_i-x_c)
\end{equation}

Note that although the expression in equation~(\ref{prachigh}) is
reminiscent of equation~(\ref{decomp}), the factor attributed to
$A(\theta)$ in equation~(\ref{prachigh}), is a function of the data
($I_i, x_i$), and therefore does not fulfill the decomposition in
equation~(\ref{decomp}). Consequently, even under the high resolution
approximation, there is no estimator that achieves the
\crra\ bound. This situation supports and justifies the adoption of
            {\it alternative criteria} for position estimation,
            maximum likelihood and the classical least squares being
            two of the more commonly used approaches adopted. These
            are reviewed in the following subsections.


\subsection{Maximum Likelihood} \label{ml}

Given the flux values $I_1,...,I_n$, the maximum-likelihood (ML)
estimate of the position $x_c$ is obtained through the following rule:

\begin{equation}\label{ml1}
\hat{x}_{c_{ML}}(I_1,...,I_n) = \arg \max_{x_c} \ln L(I_1,...,I_n;x_c)
\end{equation}

where ``$\arg \max$'' represents the argument that maximizes the
expression. Imposing the first order condition on this optimization
problem, it reduces to satisfying the condition $\frac{d \ln
  L(I_1,...,I_n;x_c)}{d x_c} = 0$, and, consequently, we can work with
the general expression in equation~(\ref{derilike1}).
We note, from that equation, that the term in the brackets given by
$\sum_{i=1}^n \lambda_i(x_c) \equiv \tilde{F} + \tilde{B}$ is
independent of $x_c$, and consequently its derivative is
zero. Therefore the ML condition becomes:

\begin{equation}\label{eq_app_CR_1D_1_b}
\frac{d \ln L(I_1,..,I_n;x_c)}{d x_c} \equiv \sum_{i=1}^n I_i \cdot
\frac{1}{\lambda_i(x_c)} \cdot \frac{d \lambda_i(x_c)}{d x_c} =
\frac{1}{\sigma^2} \sum_{i=1}^n I_i \cdot
\frac{\tilde{F}_i(x_c)}{\tilde{F}_i(x_c) + \tilde{B}_i(x_c)} \cdot
(x_i - x_c) = 0 \label{ml2}
\end{equation}

where we have used the high-resolution approximation,
equation~(\ref{derivap}). In general, this condition does not offer a
closed-form solution (note that the dependency of $\tilde{F}_i$ on
$x_c$ could be quite complex), consequently numerical gradient-descent
methods need to be implemented. Nevertheless, in the signal-dominated
regime, when $\tilde{F}_i \gg \tilde{B}_i$, it is straightforward to
show that equation~(\ref{ml2}) reduces to the classical moment
solution, i.e.:

\begin{equation} \label{ml3}
\hat{x}_{c_{ML}} = \frac{\displaystyle \sum_{i=1}^{n} I_i \cdot
  x_i}{\displaystyle \sum_{i=1}^{n} I_i }
\end{equation}

In general, the formal solution to equation~(\ref{ml2}) in the
high-resolution regime offers a simple relationship to implement a
recursive algorithm that satisfies the ML estimate, given by:

\begin{equation} \label{ml3}
\hat{x}_{c_{ML}} = \frac{\displaystyle \sum_{i=1}^{n} w_i(x_c) \cdot
  I_i \cdot x_i}{\displaystyle \sum_{i=1}^{n} w_i(x_c) \cdot I_i }
\end{equation}

where the weights $w_i(x_c) \equiv
\frac{\tilde{F}_i(x_c)}{\tilde{F}_i(x_c) + \tilde {B}_i}$.

To conclude this subsection, we must mention that it is well-known
that in the setting of independent and identically distributed
measurements, the ML estimate is asymptotically unbiased and efficient
\citep[Chap. 18]{stu04}.
In other words, as the number of samples increases, the variance tends
to the \crra\ bound and consequently the ML is asymptotically optimal
(in the sense of achieving the minimum variance bound).  However, the
astrometry setting is different as the measurements (fluxes) follow a
Poisson distribution with parameters that are position dependent (see
Section \ref{subsec:post_est_astrometry}), and consequently the ML is
not necessarily efficient in this statistical sense.

\subsection{Least Squares}\label{ls}

Given the flux values $I_1,...,I_n$, the (weighted) least square (LS)
estimate of the position $x_c$ is given by the following
decision-rule:

\begin{equation}\label{ls1}
\hat{x}_{c_{LS}}(I_1,...,I_n) = \arg \min_{x_c} \sum_{i=1}^n
\frac{\left( I_i-a_i(x_c) \right) ^2}{b_i(x_c)}
\end{equation}

being $a_i(x_c) \equiv \mathbb{E}(I_i) = \lambda_i(x_c)$ and $b_i(x_c)
\equiv Var(I_i) = \lambda_i(x_c)$ for all $i$.\footnote{Note that the
  ML reduces to the LS estimate when the data follows a Gaussian
  distribution.} The unweighted LS estimate assumes instead that
$b_i=1$, where no variance normalization is considered at all.

If we define $E_{LS}(I_1,...,I_n;x_c) \equiv \sum_{i=1}^n \left( I_i-
\lambda_i(x_c) \right)^2$, then the first-order condition over the
unweighted LS estimate implies that:

\begin{equation}\label{ls2}
\frac{d E_{LS}(I_1,...,I_n;x_c)}{d x_c} = \sum_{i=1}^n (I_i -
\lambda_i(x_c)) \cdot \frac{d \lambda_i(x_c)}{d x_c}=0
\end{equation}

Hence, using again the high-resolution regime
(equation~(\ref{derivap})), this condition implies that:

\begin{equation}\label{ls3}
\frac{d E_{LS}(I_1,...,I_n;x_c)}{d x_c} = \sum_{i=1}^{n}
\frac{\tilde{F_i}}{\sigma^2} \cdot \left( I_i - \lambda_i \right) \cdot
(x_i-x_c) = 0
\end{equation}

and, consequently, the problem reduces to solve the equation:

\begin{equation} \label{ls4}
\hat{x}_{c_{LS}} = \frac{\displaystyle \sum_{i=1}^{n} \tilde{F}_i
  \cdot \left( I_i - \lambda_i(x_c) \right) \cdot x_i}{\displaystyle
  \sum_{i=1}^{n} \tilde{F}_i \cdot \left(I_i - \lambda_i(x_c) \right)}
\end{equation}

Note that this expression is the counterpart of equation~(\ref{ml3})
for the ML estimate, emphasizing that these two techniques offer {\it
  different} estimates of the position in the parametric setting of
astrometry.  Again the solution of (\ref{ls4}) involves the use of
numerical methods, as, in general, no closed-form solution can be
derived from it.

\subsection{Numerical results} \label{numres}

In this section we present some results from numerical experiments for
the standard deviation of the astrometric position of a 1-D Gaussian
source sampled by a linear detector. The goal of these experiments is
to compare the performance of the ML and LS solutions described in the
previous subsections, to the theoretical \crra\ bound, under different
assumptions regarding the detector and the source.

Basically, we start by adopting a set of values for the gain and
read-out noise of the detector, as well as its pixel size and number
of pixels. For a source with a Gaussian PSF, we specify its width, the
maximum flux at the central pixel and its center location ($x_c$). For
the background, we adopt a certain (fixed) value per pixel. The total
flux of the source (in ADUs) is obtained through equation~(\ref{fmax})
by direct integration of the Gaussian PSF given the adopted values for
$\Delta x$, $x_c$ and $\sigma$. We generate many possible
``observations'' for the same combination of parameters, using a
random-number generator driven by a Poisson distribution using the
$poidev$, $gammaln$, and $ran1$ routines explained in
\citet{pres92}. We note that we transform all the ADUs (source and
background) to units of $e^-$ using the adopted gain, before
randomizing the data. When generating the data, we also consider the
read-out noise and the ``digitization noise'' (see, e.g,
\citet{gill92}). On output we have an array of pixel positions (1 to
$n$) and their corresponding fluxes.

After generating a (large) number of simulations for a given set of
parameters, we compute the value of $x_c$ using a ML and a LS
(weighted and unweighted) procedure. Because we are operating in the
1-D case, where we are estimating the object's position exclusively,
all the other parameters were assumed to be known, and fed to the
routine that searched for the value of $x_c$, using either
equation~(\ref{ml1}) in the case of ML or equation~({\ref{ls1}) in the
  case of LS. To estimate the optimum number of simulations required
  to obtain a stable solution, we computed a very large number of
  simulations for our first set, and then calculated the mean of the
  recovered $x_c$ and its standard deviation, $\sigma_{x_c}$, as a
  function of the number of simulations. The optimum value for the
  number of simulations should render a set of values ($x_c$,
  $\sigma_{x_c}$) that do not depend appreciably (within their
  statistical uncertainty) on adding more simulations. For our
  simulations, this number turned out to be $\sim$250 simulations per
  set.

In Table~\ref{tabnum} we present the standard deviation from the
simulations, $\sigma_{x_c}$, for a number of representative cases, as
well as the calculated \crra\ minimum variance bound based on
equation~(\ref{exact}), denoted by $\sigma_{\tiny \mbox{CR}}$.
In all cases we adopted an array size of 100~pixels that properly
covers the PSF, a pixel size of 0.2~arcsec, a detector $RON=5$~$e^-$,
a (fixed) sky background of 300~ADU per pixel, and a Gaussian source
with a $FWHM=1$~arcsec.

To our surprise, the results on Table~\ref{tabnum} indicate that the
performance of the ML and LS estimators are almost equal to the
\crra\ lower bound. This is a remarkable result, and it validates the
predictive power of the \crra\ bound and its use as a benchmark
indicator in astrometry. From the theoretical side, we formally showed
(Section~\ref{sec_pract_estimators}, equation~(\ref{prachigh})) that
the \crra\ bound for astrometry can not be achieved. Our numerical
results suggest then that, both, the ML and the LS methods (which are
widely used in astrometry), are very efficient in the sense of
asymptotically (with the number of measurements) approaching the
\crra\ bound. Proving this conjecture is an interesting topic of
further work which will help us to explain the results obtained, and
further consolidate the use the \crra\ bound for performance
analysis. A possible explanation, valid in the 1-D case, has been
proposed in Section~\ref{interp}: This explanation however is not
valid in a general situation, where one needs to simultaneously
estimate astrometric and photometric parameters, the subject of which
will be further explored in a forthcoming paper. We also note from
Table~\ref{tabnum} that the ML is as good as, or even better in some
cases, than the LS and WLS estimators.

We finally compare the predictions of the \crra\ lower bound to the
performance of some 2-D digital centering algorithms applied to
simulated stellar images, reported by \citet{sto89}. As argued in
Section~\ref{2d} we should expect that our results from the 1-D case
should not significantly differ from those of a 2-D case when the
source is symmetrical. Indeed, \citet{sto89} adopted for his
simulations symmetrical Gaussian sources (his equation~(1)) on a flat
background. We have read approximately from his Figures~2 to~5 the
results for the {\it minimum} rms (over different profile fitting
methods) of the positional uncertainty from his 2-D numerical
simulations for some representative values of the total flux $F$, this
are tabulated in our Table~\ref{tabsto} under the label
$\sigma_{\mbox{\tiny St}}$, in units or arcsec. For each of these
values we have computed the predicted \crra\ bound using our
equation~(\ref{exact}), and adopting the same values for $F$, $B$,
$FWHM$, and $\Delta x$ used by \citet{sto89} for each of his
simulations. These values are denoted by $\sigma_{\mbox{\tiny CR}}$ in
Table~\ref{tabsto}, also in units or arcsec. As can be seen from this
Table, our predictions are extremely encouraging: In all cases our
computed \crra\ bound is smaller than (although typically close to)
the results from the ``measured'' standard deviation of the position
(derived from the 2-D simulations). Furthermore, as indicated in
Section~\ref{ml} (specially equation~(\ref{ml3})), when $F>>B$ we
would expect that the ML estimate approaches the moment solution,
which is indeed the case as seen from Table~\ref{tabsto} (see also
Figures~2 and 3 for large ``counts in image'' in \citet{sto89}).

\section{Conclusions}

Our results indicate that we have found in the \crra\ lower variance
bound a very powerful astrometric ``benchmark'' estimator concerning
the maximum expected positional precision for a point source, given a
prescription for the source, the background, the detector
characteristics, and the detection process.

We regard as particularly interesting, pedagogical, and as a ``back of
an envelope'' estimation tool, the set of equations~(\ref{weakstrong})
which, albeit derived in the high-resolution regime, are actually
quite resilient to this condition, and thus provide reasonable
expectations for the astrometric precision.

In a forthcoming paper we will formally extend our analysis to the 2-D
case for the estimation of $(x,y)$ coordinates in a more realistic
pixel array. It has been argued however (see Sections~\ref{2d} and
\ref{numres}) that, as long as the PSF is reasonably symmetric, the
results on equation~(\ref{weakstrong}) are not likely to change much
in this case. In particular, the result that for background-dominated
sources the \crra\ lower bound goes as $B/F^2$, while when the
background is negligible, this maximum achievable precision goes as
$F^{-1}$ should still hold in the 2-D case.

Also, it would be interesting to study the sensitivity of the
\crra\ bound to other PSF shapes. We note however that the results by
\citet{kin83}, which coincide with our own results (see
Section~\ref{weastr}), have been derived from a different PSF (his
equation~(10)). \citet{kin83} argues that, by using a Gaussian
instead, only minor differences in the numerical coefficients in his
equations~(23) and~(24) appear.

Other factors to consider in a more realistic application include the
existence of a strong gradient in the background under the object's
PSF, which will tend to bias the derived astrometry, and possibly also
affect the \crra\ estimate (throughout this paper we have assumed that
the background is uniform, but an extension to a highly variable
background is straightforward to implement, starting from
equation~(\ref{var2})). Also, in the case of severely under-sampled
images, the issue of intra-pixel response function has been
demonstrated to be significant (see, e.g., Figure~1 in \citet{ad96}),
and should be included in the analysis.

Eventually, one would like to be able to compute the \crra\ bound in
cases where there is a simultaneous joint estimation of the
photometric (source and background), and astrometric (position, width
of the PSF) parameters, which would for sure require non-linear,
possibly iterative, numerical methods, in which careful attention
should be given to numerical stability issues, and proper handling of
observational errors - this constitutes the long-term goal of our
research.

\clearpage

\acknowledgments

RAM acknowledges partial support from the ``Center of Excellence in
Astrophysics and Associated Technologies'' (PFB 06) - CONICYT. JFS
acknowledges support from FONDECYT - CONICYT grant \#1110145. The
authors would also like to acknowledge Prof. William F. van Altena for
many useful comments on an earlier version of this paper, and an
anonymous referee for his/her many corrections and suggestions of
style, and for highlighting some aspects of our work that led to
Section~\ref{dith}.

\clearpage

\appendix


\section{Derivation of the Fisher information about $x_c$} \label{app_fisher}


We start from the expression in equation~(\ref{derilike1}):

\begin{equation}\label{eq_app_CR_1D_1_b}
\frac{d \ln L(I_1,..,I_n;x_c)}{d x_c}  =\sum_{i=1}^n I_i \cdot
\frac{1}{\lambda_i(x_c)} \cdot \frac{d \lambda_i(x_c)}{d x_c}
\end{equation}

where, without any loss of generality, we have assumed that
$\sum_{i=1}^n \lambda_i(x_c)$ is constant, independent of the
parameter $x_c$ (see details on Section~\ref{ml}).  Then, the Fisher
information about $x_c$ can be obtained as follows (see
equation~(\ref{fisher})):

\begin{eqnarray}\label{eq_app_CR_1D_2_b}
\mathcal{I}_{x_c}(n) & \equiv \mathbb{E}_{I_1,...,I_n} \left( \left(
\frac{d }{d x_c} \ln L(I_1,...,I_n;x_c) \right)^2 \right) \nonumber\\
 & = \mathbb{E}_{I_1,...,I_n} \left( \sum_{i=1}^n \sum_{j=1}^n
\frac{I_i \cdot I_j}{\lambda_i(x_c) \cdot \lambda_j(x_c)} \cdot \frac{d
  \lambda_i(x_c)}{d x_c} \cdot \frac{d \lambda_j(x_c)}{d x_c} \right)
\nonumber\\
 & = \mathbb{E}_{I_1,...,I_n} \left( \sum_{i} \left(\frac{I_i
}{\lambda_i(x_c)} \cdot \frac{d \lambda_i(x_c)}{d x_c} \right)^2
\right) + \mathbb{E}_{I_1,...,I_n} \left( \sum_{i}\sum_{j \neq i}
\frac{I_i \cdot I_j }{\lambda_i(x_c) \cdot \lambda_j(x_c)} \cdot \frac{d
  \lambda_i(x_c)}{d x_c} \cdot \frac{d \lambda_j(x_c)}{d x_c}
\right)\nonumber\\
 & = \sum_{i} \frac{\left( \lambda_i(x_c)+
  \lambda_i(x_c)^2 \right)}{\lambda_i(x_c)^2} \cdot \left( \frac{d \lambda_i(x_c)}{d
  x_c} \right)^2 + \sum_{i}\sum_{j \neq i} \frac{d \lambda_i(x_c)}{d
  x_c} \cdot \frac{d \lambda_j(x_c)}{d x_c}\nonumber\\ 
& = \sum_{i}
\frac{1}{\lambda_i(x_c)} \cdot \left( \frac{d \lambda_i(x_c)}{d x_c}
\right)^2 + \sum_{i} \left( \frac{d \lambda_i(x_c)}{d x_c} \right)^2 +
\sum_{i}\sum_{j \neq i} \frac{d \lambda_i(x_c)}{d x_c} \cdot \frac{d
  \lambda_j(x_c)}{d x_c} \nonumber\\
 & = \sum_{i} \frac{1}{\lambda_i(x_c)} \cdot \left( \frac{d
  \lambda_i(x_c)}{d x_c} \right)^2 + \left( \sum_{i} \frac{d
  \lambda_i(x_c)}{d x_c} \right)^2 \nonumber\\
 & = \sum_{i} \frac{1}{\lambda_i(x_c)} \cdot \left( \frac{d
  \lambda_i(x_c)}{d x_c} \right)^2 + \left( \frac{d}{d x_c}
\sum_{i=1}^n \lambda_i(x_c) \right)^2 = \sum_{i=1}^n
\frac{1}{\lambda_i(x_c)} \cdot \left( \frac{d \lambda_i(x_c)}{d x_c}
\right)^2
\end{eqnarray}

where we have used the fact that, for a Poisson distribution, $
\lambda_i = \mathbb{E}(Var(I_i)) \equiv \mathbb{E} \left(I_i-\lambda_i
\right)^2 = \mathbb{E}(I_i^2) - \lambda_i^2$.

\clearpage

\begin{figure}
\epsscale{.80}
\plotone{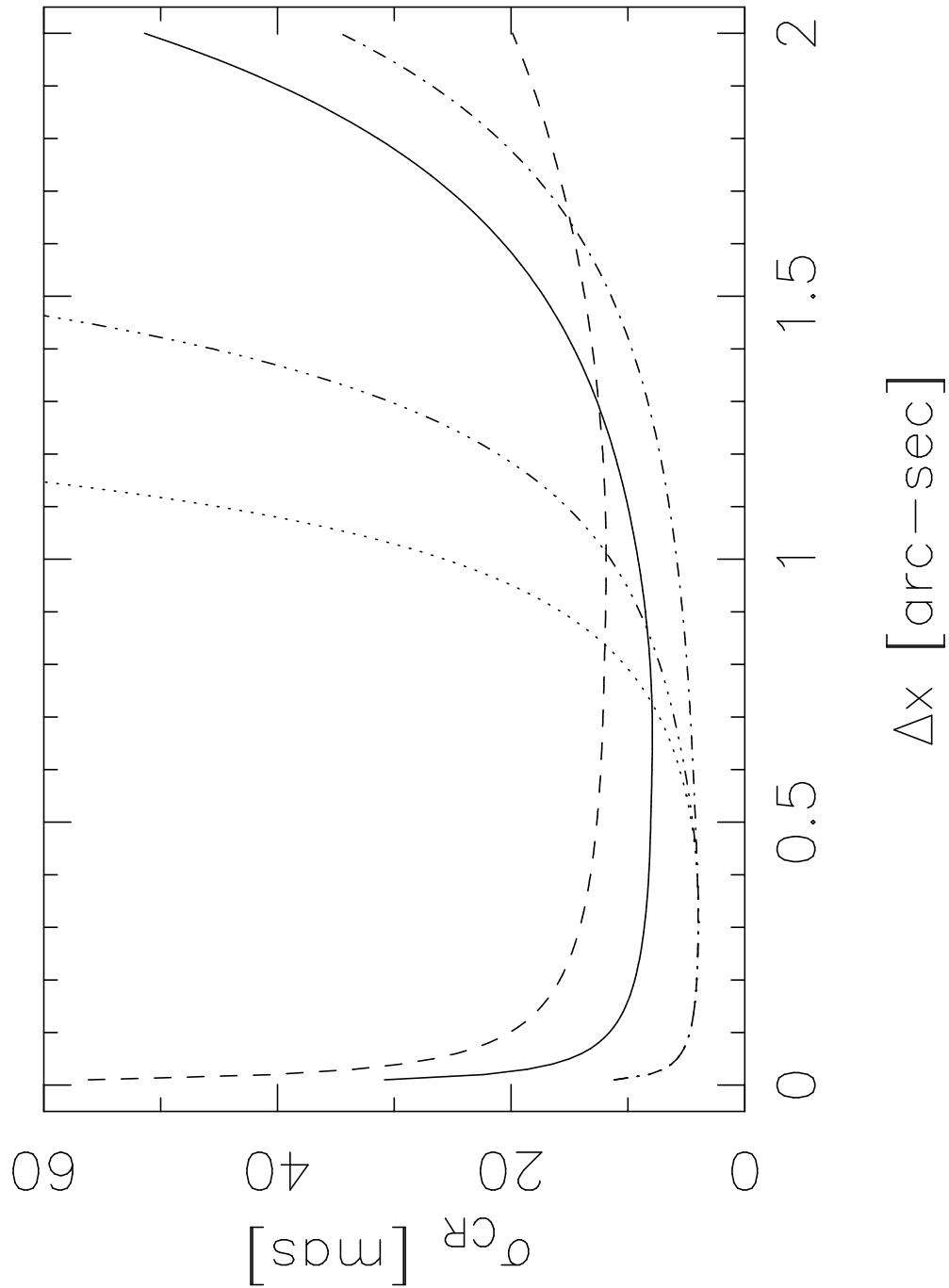}
\caption{\crra\ bound as given be equation~(\ref{exact}), in
  milli-arcsec (mas), as a function of detector pixel size $\Delta x$ in
  arcsec. All curves were computed for a constant background per
  pixel $B=300$~ADU/pix and $F/B = 10$. The dotted, solid, and dashed
  lines are for a Gaussian PSF with a $FWHM$ of 0.5, 1.0, and
  1.5~arcsec respectively, all centered in a given pixel of width 1.0
  pixel (compare to Figure~1 in \citet{win86}). The dashed-triple
  dotted and dashed-dotted lines are for a $FWHM$ of 0.5 arcsec, but
  de-centered by 0.125~pix and 0.25~pix from the center of the pixel
  respectively. \label{figcr1}}
\end{figure}

\clearpage

\begin{figure}
\epsscale{.80}
\plotone{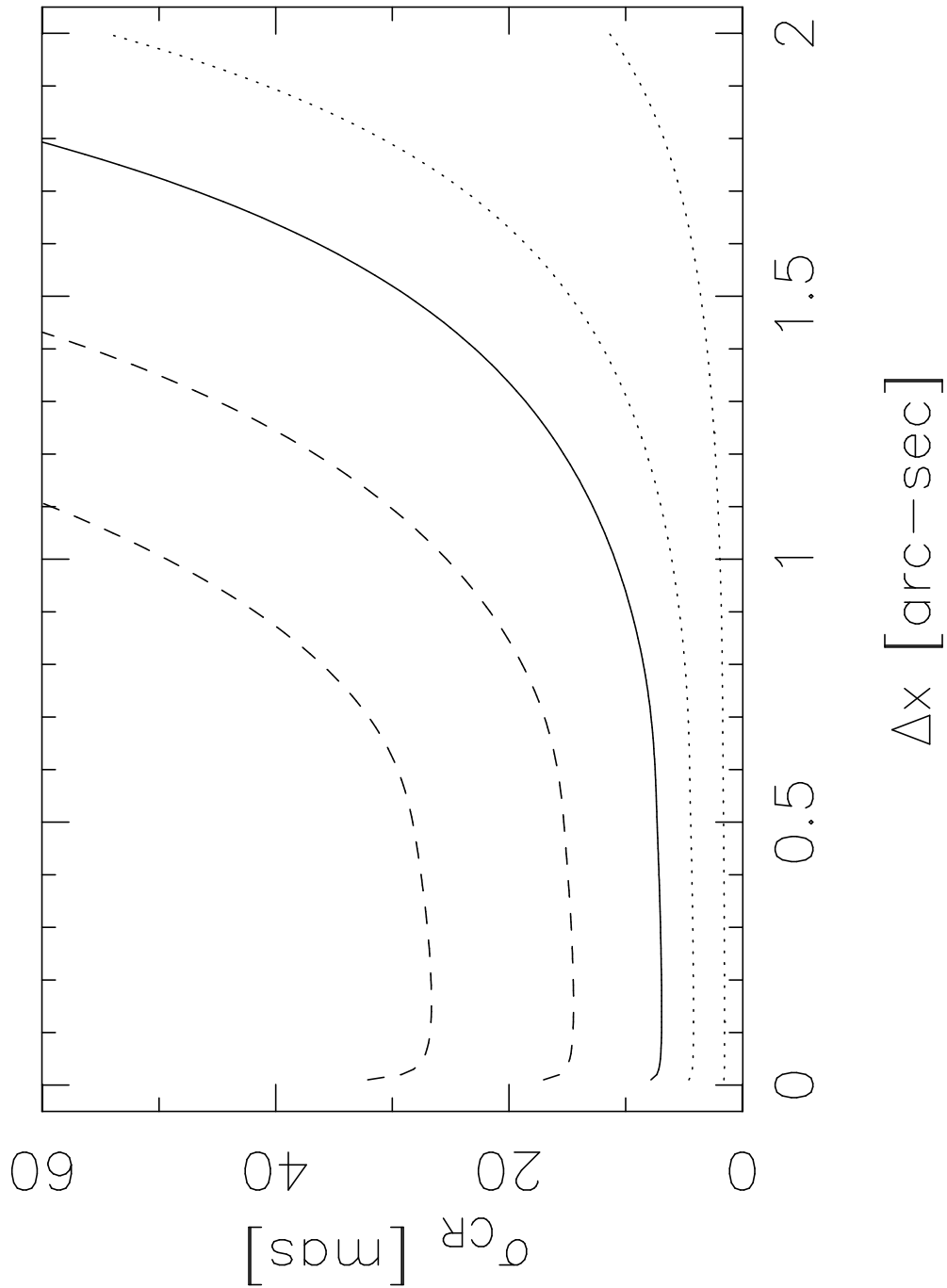}
\caption{\crra\ bound as given be equation~(\ref{exact}), in
  milli-arcsec (mas), as a function of pixel size $\Delta x$ in
  arcsec. All curves were computed for a background given by
  equation~(\ref{back}) with $f_s=2\,000$~ADU/arcsec, $RON=5$~e$^-$,
  $D=0$~e$^-$, $G=2$~e$^-$/ADU, and for a Gaussian source with
  $FWHM=1$~arcsec centered on a pixel. The curves shown have different
  values of the flux, and hence a different $S/N$. From top to bottom
  we have $F=1\,000$~ADU, $S/N \approx 20$, $F=2\,000$~ADU, $S/N
  \approx 35$ (both dashed lines); $F=5\,000$~ADU, $S/N \approx 74$
  (solid line, as a reference, in this case we have $F_{\mbox{max}}
  \approx 930$~ADU at $\Delta x=0.2$~arcsec, see
  equation~(\ref{fluxapp2})); and $F=10\,000$~ADU, $S/N \approx 120$,
  and $F=50\,000$~ADU, $S/N \approx 300$ (both dotted
  lines). \label{figcr2}}
\end{figure}

\clearpage

\begin{table}
\begin{center}
\caption{\crra\ bound on as a function of pixel offset for a source
  with $FWHM=0.5$~arcsec observed with two detectors of pixel size
  indicated by the $\Delta x$ value. Other parameters are those of
  Fig.~\ref{figcr1}.\label{tabdit}}
\begin{tabular}{ccr}
\tableline\tableline
$\Delta x$ & Pixel offset & $\sigma_{\mbox{\tiny CR}}$ \\
arcsec & pix & mas \\
\tableline
0.5 & 0.0 & 4.57 \\
0.5 & 0.125 & 4.46 \\
0.5 & 0.250 & 4.22 \\
\tableline
0.7 & 0.0 & 7.31 \\
0.7 & 0.125 & 6.06 \\
0.7 & 0.250 & 4.65 \\
\tableline
0.9 & 0.0 & 15.70 \\
0.9 & 0.125 & 9.00 \\
0.9 & 0.250 & 5.33 \\
\tableline
\tableline
\end{tabular}
\end{center}
\end{table}

\clearpage

\begin{table}
\begin{center}
\caption{Small-pixel approximation and exact \crra\ bound on some
  limiting cases.\label{tabcomp1}}
\begin{tabular}{cccccccc}
\tableline\tableline
$\Delta x$ & $FWHM$ & $F$ & $F_{\mbox{max}}$ & $f_s$ &
$S/N$\tablenotemark{a} & Exact\tablenotemark{b} &
Approx\tablenotemark{c} \\
arcsec & arcsec & ADU& ADU & ADU & & mas & mas \\
\tableline
0.2 & 1.0 & $ 1\,000$ & $\sim$186    & $2\,000$ & $\sim$20  & 27   & 24 \\
0.2 & 1.0 & $50\,000$ & $\sim9\,308$ & $2\,000$ & $\sim$300 & 1.5  & 1.4 \\
0.2 & 1.0 & $50\,000$ & $\sim9\,308$ & $500$    & $\sim$305 & 1.4  & 1.4 \\
\tableline
0.1 & 0.5 & $50\,000$ & $\sim9\,308$ & $4\,000$ & $\sim$300 & 0.76 & 0.67 \\
0.1 & 0.5 & $1\,000$  & $\sim186$    & $4\,000$ & $\sim$20  & 13   & 12 \\
\tableline
0.1 & 0.5 & $5\,000$  & $\sim1\,024$ & $4\,000$ & $\sim$78  & 3.5  & 2.4/2.1\tablenotemark{d} \\
0.2 & 1.0 & $5\,000$  & $\sim1\,024$ & $2\,000$ & $\sim$78  & 6.9  & 4.8/4.3\tablenotemark{d} \\
\tableline
\tableline
\end{tabular}

\tablenotetext{a}{As computed from equation~\ref{sn3} with
  $RON=5$~e$^-$, $G=2$~e$^-$/ADU, and the tabular value for
  $\Delta x$ and $f_s$.}

\tablenotetext{b}{``Exact'' \crra\ bound
  computed from equation \ref{exact}.}

\tablenotetext{c}{Small-pixel approximation \crra\ bound computed from
  equation \ref{weakstrong}, depending on if $S/N$ is large or small.}

\tablenotetext{d}{The first value is for the case of a weak source,
  the second for a strong source, both from equation
  \ref{weakstrong}.}

\end{center}
\end{table}

\clearpage

\begin{table}
\begin{center}
\caption{Comparison of the standard deviation on position
  $\sigma_{x_c}$ from the ML and LS methods and the \crra\ bound,
  $\sigma_{\tiny \mbox{CR}}$, as a function of (total)
  Flux.\label{tabnum}}
\begin{tabular}{cccccccc}
\tableline\tableline
&&& \small{Method:} & ML & LS & WLS &  Cram\'er-Rao \\
\tableline
$F$ & $F_{\small\mbox{max}}$ & $S/N$ & & $\sigma_{x_c}$ & $\sigma_{x_c}$ & $\sigma_{x_c}$ &  $\sigma_{\tiny \mbox{CR}}$ \\
ADU & ADU & & & pix & pix & pix & pix \\
\tableline
$30,080$ & 5600 & $\sim230$  & & $0.0101 \pm 0.0005$ & $0.0116 \pm 0.0006$  & $0.0103 \pm 0.0005$ & 0.010 \\
$10,002$ & 1862 & $\sim120$  & & $0.0215 \pm 0.0010$ & $0.0227 \pm 0.0010$  & $0.0215 \pm 0.0010$ & 0.020 \\
$3,222$  &  600 & $\sim55$   & & $0.0429 \pm 0.0019$ & $0.0437 \pm 0.0020$  & $0.0431 \pm 0.0019$ & 0.045 \\
$1,612$  &  300 & $\sim32$   & & $0.080 \pm 0.003$   & $0.081 \pm 0.004$    & $0.080 \pm 0.003$   & 0.079 \\
$540$    &  100 & $\sim12$   & & $0.212 \pm 0.010$   & $0.212 \pm 0.010$    & $0.212 \pm 0.010$   & 0.209 \\
$268$    &   50 & $\sim6$    & & $0.423 \pm 0.020$   & $0.423 \pm 0.020$    & $0.423 \pm 0.020$   & 0.406 \\
\tableline
\tableline
\end{tabular}
\end{center}
\tablenotetext{a}{The $1\sigma$ uncertainties in the derived standard
  deviation were computed from the variance of the variance,
  $Var(\sigma^2_{x_c})$, as given by equation~(3.44) in
  \citet{roe10}. It is easy to show that the uncertainty in the
  standard deviation would be then given by
  $\sqrt{Var(\sigma^2_{x_c})}/2 \, \sigma_{x_c}$.}
\end{table}

\clearpage

\begin{deluxetable}{c|cc|cc|cc|cc}
\tabletypesize{\scriptsize}
\rotate
\tablecaption{Comparison of the best 2-d digital centering algorithms (from
  \citet{sto89}) and the \crra\ bound as a function of (total)
  Flux.\label{tabsto}}
\tablewidth{0pt}
\tablehead{
\multicolumn{1}{c}{} & 
\multicolumn{2}{c}{$FWHM$=1~arcsec, $B$=1~ADU/arcsec}    &
\multicolumn{2}{c}{$FWHM$=4~arcsec, $B$=1~ADU/arcsec}    &
\multicolumn{2}{c}{$FWHM$=1~arcsec, $B$=1000~ADU/arcsec} &
\multicolumn{2}{c}{$FWHM$=4~arcsec, $B$=1000~ADU/arcsec} \\
\multicolumn{1}{c}{} & 
\multicolumn{2}{c}{$\Delta x$ = 0.125~arcsec} &
\multicolumn{2}{c}{$\Delta x$ = 0.5  ~arcsec} &
\multicolumn{2}{c}{$\Delta x$ = 0.125~arcsec} &
\multicolumn{2}{c}{$\Delta x$ = 0.5  ~arcsec} \\
F & $\sigma_{\mbox{\tiny St}}$ & $\sigma_{\mbox{\tiny CR}}$ & $\sigma_{\mbox{\tiny St}}$
& $\sigma_{\mbox{\tiny CR}}$ & $\sigma_{\mbox{\tiny St}}$ & $\sigma_{\mbox{\tiny CR}}$ &
$\sigma_{\mbox{\tiny St}}$ & $\sigma_{\mbox{\tiny CR}}$ \\
ADU & arcsec & arcsec & arcsec & arcsec & arcsec & arcsec & arcsec & arcsec \\
}
\startdata
$100$      & 0.052 & 0.047 & 0.4 & 0.2 & --- & 0.24 & --- & 1.9 \\
$1,000$    & 0.015 & 0.014 & 0.07 & 0.055 & 0.04 & 0.029 & --- & 0.20 \\
$10,000$   & 0.042\tablenotemark{a} & 0.043 & 0.02 & 0.017 & 0.006 & 0.0053 & 0.07\tablenotemark{c} & 0.027 \\
$100,000$  & 0.0012\tablenotemark{a} & 0.0014 & 0.005\tablenotemark{a} & 0.0054 & 0.0015\tablenotemark{b} & 0.0014 & 0.009\tablenotemark{c} & 0.0060 \\
\enddata
\tablenotetext{a}{Moment solution.}
\tablenotetext{b}{Far from moment solution, as expected since $B$ is large.}
\tablenotetext{c}{No moment solution found.}
\tablecomments{All \crra\ estimates used $G=1$~e$^-$/ADU and $RON=0$~e$^-$.}
\end{deluxetable}

\end{document}